%% file: main.tex
\documentclass[showpacs,aps,prc,longbibliography,showkeys,superscriptaddress,twocolumn]{revtex4-1}

\usepackage[colorlinks, urlcolor=blue,linkcolor=blue, anchorcolor=blue, citecolor=blue]{hyperref}
\usepackage[english]{babel}
\usepackage[utf8]{inputenc}
\usepackage{multirow}
\usepackage{array}
\usepackage{amsthm}
\usepackage{mathtools}
\usepackage{physics}
\usepackage{xcolor}
\usepackage{graphicx}
\usepackage{bm}
\usepackage{soul}
\usepackage{adjustbox}
\usepackage{placeins}
\usepackage[T1]{fontenc}
\usepackage{lipsum}
\usepackage{csquotes}
\usepackage{float}
\usepackage{booktabs}
\usepackage{lineno}

\graphicspath{{figs/}}

\newcommand{\llbar}{\ell\bar\ell}

\newcommand{\Appendix}{Appendix}

\begin{document}

\title{Origin and limits of intermediate-mass dilepton thermometry}

\author{Lipei Du}
\affiliation{Department of Physics, University of California, Berkeley CA 94720}
\affiliation{Nuclear Science Division, Lawrence Berkeley National Laboratory, Berkeley CA 94720}

\date{\today}

\input{body}

\section*{Acknowledgements}
The author acknowledges helpful conversations with Charles Gale and Jean-Fran\c{c}ois Paquet. This work was supported in part by the U.S. Department of Energy, Office of Science,
Office of Nuclear Physics, under Grant No.~DE-AC02-05CH11231. The author acknowledges the use of ChatGPT for grammar refinement, clarity enhancement, and analysis code optimization.

\bibliography{refs}


\appendix*

\renewcommand{\thefigure}{S\arabic{figure}}
\setcounter{figure}{0} 
\setcounter{equation}{0} 
\renewcommand{\thetable}{\arabic{table}}


\input{app}
\end{document}

%% file: body.tex
\begin{abstract}
We identify the physical origin and limits of intermediate-mass dilepton
thermometry in relativistic heavy-ion collisions. Using a controlled
expanding-fireball framework with thermal dilepton rates, we show that the
local inverse-slope parameter of the invariant-mass spectrum follows, to
percent-level accuracy, the emission-weighted harmonic mean of the
temperatures contributing to the spectrum. As the thermal-stage initial
temperature increases, the temperatures sampled by the radiation shift upward
with the overall thermal scale, causing their harmonic mean to track the
initial temperature closely. This provides the physical basis for the strong
inverse-slope--initial-temperature correlation, while the resulting mapping
remains nonuniversal: changes that only rescale the amount of radiation leave
it unchanged, whereas changes in the cooling history or source composition
redistribute the radiation among different temperatures and modify the
response. The mapping is nevertheless nearly linear over the temperature
range studied here. Because different invariant-mass windows weight the
emission history differently, source scenarios with the same inverse-slope
parameter in $1<M<3~{\rm GeV}$ develop different inverse slopes in harder
mass windows. Intermediate-mass dilepton spectra therefore provide a
quantitative but nonuniversal probe of the early thermal history, while
measurements in multiple mass windows, when confronted with realistic
calculations, can provide additional constraints on the thermal evolution and
early electromagnetic source content.
\end{abstract}

\maketitle

\section{Introduction}
Determining the temperature reached in relativistic heavy-ion collisions is
essential for characterizing the quark-gluon plasma (QGP) and its space-time
evolution~\cite{Shuryak:2014zxa,Heinz:2013th}. Final-state hadrons are strongly
shaped by the later evolution and freeze-out, after the medium has cooled and
developed substantial collective flow~\cite{Arslandok:2023utm}.
Electromagnetic radiation instead escapes with little final-state interaction
and carries information from all stages of the collision
\cite{Peitzmann:2001mz,Gale:2009gc,Rapp2013,Linnyk2015,Salabura:2020tou}.
Its interpretation is nevertheless nontrivial because an observed spectrum
superposes radiation emitted over a broad range of times, temperatures, and
microscopic sources~\cite{Geurts:2022xmk}.

Dileptons in the intermediate-mass region (IMR) provide a particularly
promising temperature probe. Their invariant mass is not blue-shifted by
collective flow in the same manner as transverse-momentum spectra, while the
Boltzmann suppression at large mass preferentially selects radiation from
hotter stages of the evolution
\cite{Shuryak:1978ij,Kajantie:1981wg,Kajantie:1986dh,McLerran1984}. After
removal of the leading mass prefactor, the spectrum is approximately
exponential over a finite mass interval, allowing an effective inverse-slope
parameter $T_{\llbar}$ to be extracted
\cite{Rapp:2014hha,Churchill:2023vpt,HADES:2019auv,STAR:2024bpc}. A recent
hydrodynamic study found a nearly linear relation between this quantity and an
energy- and flow-weighted initial temperature $\langle T_{\rm in}\rangle$,
$\langle T_{\rm in}\rangle=\kappa T_{\llbar}+c$, across collision energies
and centralities~\cite{Churchill:2023zkk}. This result suggests that the IMR
mass spectrum can provide quantitative information on the early temperature
\cite{Massen:2024pnj,Du:2024pbd,Du:2025dot}.

The existence of such a thermometer relation, however, does not by itself
explain why it works. The fitted $T_{\llbar}$ is not the initial temperature,
but reflects radiation accumulated throughout the space-time evolution
\cite{Churchill:2023vpt,Vujanovic2013,Vujanovic2019}. It is therefore
important to determine what temperature scale the spectral inverse slope
actually measures, why that scale remains strongly correlated with
$\langle T_{\rm in}\rangle$, and which changes in the evolution or source
content modify this connection. These questions must be resolved before the
thermometer relation can be interpreted as more than an empirical mapping
between two temperature scales.

In this study, we identify the physical origin and principal limits of IMR
dilepton thermometry. We first show that the local inverse-slope parameter of
the mass spectrum follows, to percent-level accuracy, the emission-weighted
harmonic mean of the temperatures contributing to the radiation. We then show
that these sampled temperatures shift with the thermal-stage initial
temperature in a way that naturally produces a strong thermometer
correlation, while residual changes in their distribution make the mapping
nonuniversal. We determine how changes in cooling and source composition
modify this response and show that the mass dependence of the inverse-slope
parameter carries additional information when more than one emission source
contributes.

\section{Controlled fireball and emission setup}
To isolate the physics underlying IMR dilepton thermometry, we use a
cylindrically symmetric, boost-invariant fireball
\cite{Paquet:2023bdx,Paquet:2022wgu} rather than a full hydrodynamic
evolution. This reduced setup keeps the relevant temperature history and
emission weighting explicit, allowing individual ingredients to be varied
independently. The temperature field is parametrized by a Gaussian transverse
profile with Bjorken-like cooling,
\begin{equation}
T(\tau,r)=T_{0,{\rm peak}}
\left(\frac{\tau_0}{\tau}\right)^{c_{\rm eff}}
\exp\left[-\frac{r^2}{2\sigma_0^2}\right].
\label{eq:fireball-temperature}
\end{equation}
The baseline uses $\tau_0=1~{\rm fm}/c$ and $c_{\rm eff}=1/3$,
corresponding to conformal Bjorken cooling~\cite{Bjorken:1982qr}. The
transverse width is held fixed during the evolution, so changes of the
temperature scale can be studied without simultaneously introducing an
additional transverse-expansion timescale.

We define the thermal-stage initial temperature on the initialization plane
at $\tau_0$ using the same $e\gamma$ weighting employed in hydrodynamic
calculations,
\begin{equation}
\left\langle T_{\rm in}\right\rangle
=
\frac{
\int d^2x_\perp\,
T(\tau_0,\mathbf{x}_\perp)
e[T(\tau_0,\mathbf{x}_\perp)]
\gamma(\tau_0,\mathbf{x}_\perp)
}{
\int d^2x_\perp\,
e[T(\tau_0,\mathbf{x}_\perp)]
\gamma(\tau_0,\mathbf{x}_\perp)
},
\qquad
\gamma=u^t.
\label{eq:initial-temperature}
\end{equation}
The integrals include the initialized medium above the thermal-emission cutoff
specified below, and we use $e(T)\propto T^4$. This weighting emphasizes the
hotter part of the initial fireball and provides the reduced-model analogue
of the $e\gamma$-weighted temperature used in hydrodynamic calculations
\cite{Shen:2013vja,Churchill:2023zkk}.

The local thermal dilepton rate
\cite{Gale1987,Laine:2013vma,Ghisoiu2014,Ghiglieri2014,Jackson2019}
depends on the pair momentum through the local-rest-frame energy
$q\cdot u$~\cite{Churchill:2023vpt}. The thermal mass spectrum at midrapidity
is obtained from
\begin{equation}
\left.\frac{dN_{\rm th}}{dM\,dy}\right|_{y=0}
=
\int_{p_T} dp_T\,2\pi M p_T
\int d^4x\,
\frac{d\Gamma_{\rm th}}{d^4q}
\left(M,q\cdot u,T(x)\right).
\label{eq:thermal-dilepton-spectrum}
\end{equation}
We use $0.2<p_T<4.5~{\rm GeV}$. Because invariant mass is Lorentz invariant,
a fully momentum-integrated mass spectrum is not blue-shifted by collective
flow in the same way as a transverse-momentum spectrum
\cite{Du:2024pbd,Paquet:2022wgu,Paquet:2023bdx}. The IMR mass spectrum is
therefore much less sensitive to radial flow, apart from residual effects
associated with the finite momentum integration. We consequently neglect
transverse flow in the calculations considered here; accordingly,
$\gamma=1$ in Eq.~\eqref{eq:initial-temperature}. Thermal dilepton emission
is included only for $T\geq0.18~{\rm GeV}$, deliberately isolating the
high-temperature contribution without introducing a separate hadronic
dilepton rate.

To examine radiation emitted before $\tau_0$, we add a phenomenological
pre-equilibrium contribution extending back to
$\tau_{\rm init}=0.1~{\rm fm}/c$~\cite{Gale:2021emg}. Its effective
temperature is obtained from a backward continuation of the thermal cooling
trajectory~\cite{Paquet:2022wgu}. We use the same finite-mass dilepton rate
as the spectral input, but modify its local strength to account for the
chemically undersaturated early stage
\cite{Kurkela:2018xxd,Kurkela:2018oqw}. This effective temperature is a
variable entering the early-source prescription and should not be interpreted
as an equilibrium thermodynamic temperature. We characterize the approach
to quark chemical equilibrium by a fugacity $\lambda_q(\tau)$, with
$\tau_{\rm chem}$ defined as the time at which the quark occupancy reaches
$90\%$ of its equilibrium value~\cite{Gale:2021emg}. Since the leading IMR
process is $q\bar q\rightarrow\gamma^*\rightarrow e^+e^-$, factorized quark
and antiquark occupancies motivate weighting the local rate in
Eq.~\eqref{eq:thermal-dilepton-spectrum} by $\lambda_q^2(\tau)$ and
performing the same space-time and momentum integrations
\cite{Wu:2024pba}. This provides a controlled, Born-motivated proxy for an
early chemically undersaturated source rather than a microscopic description
of pre-equilibrium dynamics
\cite{Kurkela:2018wud,Kurkela:2018vqr,Schlichting:2019abc,
Berges:2020fwq,Coquet:2021lca}.

The relative strengths of the thermal and early contributions are fixed using
electromagnetic-source inputs constrained by realistic hydrodynamic
calculations~\cite{Churchill:2023vpt,Churchill:2023zkk} and experimental
measurements
\cite{STAR2015,STAR:2013pwb,STAR:2015zal,STAR:2024bpc,STAR:2023wta}.
We denote the resulting midrapidity spectrum, for either thermal emission
alone or the combined thermal and early contributions, by
$Y(M)\equiv
dN/(dM\,dy)|_{y=0}$.
After removing the leading asymptotic mass prefactor,
$\widetilde{Y}(M)=M^{-3/2}Y(M)$, we determine the effective inverse-slope
parameter $T_{\llbar}^{\mathcal W}$ by fitting
\begin{equation}
\ln\widetilde{Y}(M)
=
a_{\mathcal W}
-
M/T_{\llbar}^{\mathcal W}
\end{equation}
over the selected mass window $\mathcal W$. Unless stated otherwise, we use
$1<M<3~{\rm GeV}$ and suppress the superscript $\mathcal W$, writing simply
$T_{\llbar}$.

\section{Origin of the thermometer relation}

We first determine what physical temperature scale $T_{\llbar}$ measures and
then ask why it tracks $\langle T_{\rm in}\rangle$ so closely. We begin with
the baseline thermal-only calculation, using $c_{\rm eff}=1/3$ and omitting
the pre-equilibrium component. We vary $T_{0,{\rm peak}}$ from $0.24$ to
$0.48~{\rm GeV}$ while keeping all other model inputs fixed, calculate the
corresponding dilepton mass spectra, and extract $T_{\llbar}$ as defined
above. This controlled scan isolates the spectral response to changes in the
thermal-stage initial temperature.

\begin{figure}
\centering
\includegraphics[width=\linewidth]{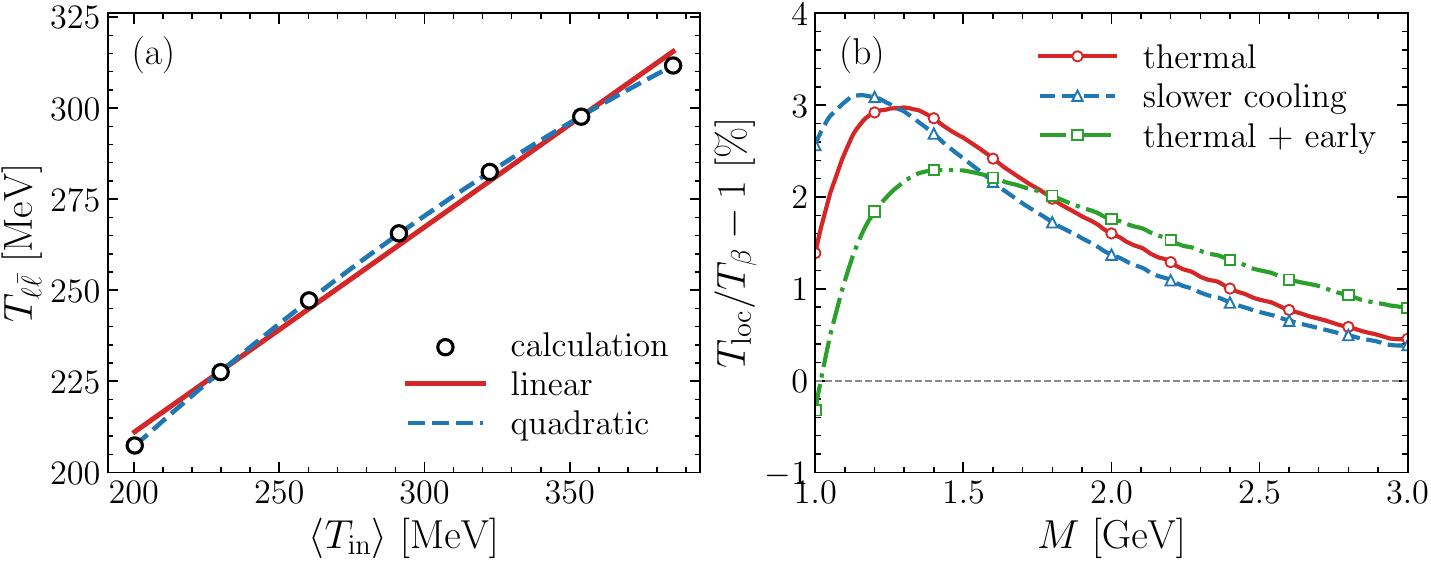}
\caption{(a) Effective inverse-slope parameter in $1<M<3~{\rm GeV}$ versus the $e\gamma$-weighted thermal-stage initial temperature when only the peak initial temperature is varied. Solid and dashed curves show linear and quadratic descriptions, respectively. (b) Fractional difference between the local inverse-slope parameter and the harmonic emission temperature for the baseline thermal evolution, slower cooling, and thermal emission supplemented by the early component.}
\label{fig:thermometer-mechanism}
\end{figure}

The resulting $T_{\llbar}$--$\langle T_{\rm in}\rangle$ relation is shown in
Fig.~\ref{fig:thermometer-mechanism}(a). We parametrize its approximately
linear response as~\cite{fnlinear}
\begin{equation}
T_{\llbar}
=
T_{\llbar}^{*}
+
B\left(
\langle T_{\rm in}\rangle-T_{\rm in}^{*}
\right),
\label{eq:linear-response}
\end{equation}
where $B$ is the response coefficient and $T_{\llbar}^{*}$ is the value of
$T_{\llbar}$ at the reference temperature $T_{\rm in}^{*}$. We refer below
to this approximately linear mapping between $T_{\llbar}$ and
$\langle T_{\rm in}\rangle$ as the thermometer relation. We choose
$T_{\rm in}^{*}=291.9~{\rm MeV}$, the mean of the sampled
$\langle T_{\rm in}\rangle$ values, so that $T_{\llbar}^{*}$ characterizes
the center of the sampled temperature range rather than an extrapolated
zero-temperature intercept. For the baseline thermal calculation,
$T_{\llbar}^{*}=262.8~{\rm MeV}$ and $B=0.564$. The maximum deviation from
the linear form is $3.85~{\rm MeV}$; including a quadratic term reduces it to
$0.18~{\rm MeV}$. The thermometer relation is therefore remarkably close to
linear over the temperature range considered here.

To understand what $T_{\llbar}$ measures, we resolve the thermal spectrum
according to the temperature at which the radiation is emitted. We introduce
$d\widetilde{Y}(M)/dT$ as the contribution to $\widetilde{Y}(M)$ from
regions of the fireball emitting near temperature $T$. To separate the
dominant Boltzmann-like mass dependence, we write
\begin{equation}
\frac{d\widetilde{Y}(M)}{dT}
\equiv
K(M,T)e^{-M/T},
\qquad
\widetilde{Y}(M)
=
\int dT\,\frac{d\widetilde{Y}(M)}{dT}.
\label{eq:kernel-representation}
\end{equation}
Thus
$K(M,T)=e^{M/T}d\widetilde{Y}(M)/dT$ contains the space-time weight,
momentum integration, and remaining mass dependence of the microscopic rate
after the leading exponential factor has been removed.
Equation~\eqref{eq:kernel-representation} is an exact rearrangement of the
calculated yield rather than an additional approximation. After removal of
the leading Boltzmann factor and asymptotic mass prefactor, the remaining
mass dependence varies more slowly across the IMR, suggesting that its
contribution to the local inverse slope should be subleading.

We then define the normalized emission-temperature distribution
\begin{equation}
P_M(T)
\equiv
\frac{1}{\widetilde{Y}(M)}
\frac{d\widetilde{Y}(M)}{dT},
\qquad
\int dT\,P_M(T)=1.
\label{eq:temperature-exposure}
\end{equation}
At fixed invariant mass, $P_M(T)\,dT$ gives the fraction of the yield emitted
from temperatures between $T$ and $T+dT$ and therefore identifies which
parts of the temperature history are sampled by the dilepton spectrum. For
any function $g(T)$, we denote the corresponding emission-weighted average by
$\langle g(T)\rangle_M
\equiv
\int dT\,P_M(T)g(T)$.
For the mixed-source calculation below, the same notation is used for the
effective early-source temperature defined above.

The connection between this temperature distribution and the spectral slope
follows by taking the logarithmic derivative of
Eq.~\eqref{eq:kernel-representation},
\begin{equation}
\frac{1}{T_{\rm loc}(M)}
\equiv
-\frac{d\ln\widetilde{Y}(M)}{dM}
=
\left\langle\frac{1}{T}\right\rangle_M
-
\left\langle
\frac{\partial\ln K(M,T)}{\partial M}
\right\rangle_M.
\label{eq:local-harmonic-relation}
\end{equation}
Here $T_{\rm loc}(M)$ is the local inverse-slope parameter of the mass
spectrum. The first term arises directly from differentiating the Boltzmann
factor and is the emission-weighted mean inverse temperature. Its inverse,
$T_\beta(M)
\equiv
\left\langle1/T\right\rangle_M^{-1}$,
with $\beta\equiv1/T$, is the emission-weighted harmonic mean of the
temperatures contributing at mass $M$, which we refer to as the harmonic
emission temperature.

The second term in Eq.~\eqref{eq:local-harmonic-relation} measures the
residual mass dependence of the emission kernel after the dominant
Boltzmann-like dependence has been separated. When this contribution is
small,
\begin{equation}
T_{\rm loc}(M)\simeq T_\beta(M),
\end{equation}
so the local inverse-slope parameter directly measures the harmonic
temperature of the emitting regions. The fitted $T_{\llbar}$ is the
corresponding finite-window characterization of this local behavior over
$1<M<3~{\rm GeV}$ for the default mass selection and therefore probes the
same underlying emission-temperature scale.

Figure~\ref{fig:thermometer-mechanism}(b) tests this interpretation directly
in three representative scenarios that will also be used below. The baseline
case contains thermal emission only with $c_{\rm eff}=1/3$. The
slower-cooling case remains thermal only but changes the cooling exponent to
$c_{\rm eff}=1/4$. The mixed-source case retains the baseline thermal
evolution and adds the chemically undersaturated early component from
$\tau_{\rm init}<\tau<\tau_0$. In all three cases, $T_{\rm loc}$ follows
$T_\beta$ to within approximately $3\%$ throughout
$1<M<3~{\rm GeV}$, with root-mean-square differences of about
$1.6$--$1.9\%$. The local IMR inverse-slope parameter therefore measures, to
percent-level accuracy, the emission-weighted harmonic mean of the
temperatures controlling the exponential mass suppression, rather than the
initial temperature itself or a simple arithmetic average over the
evolution.

The remaining question is why $T_\beta(M)$ follows $\langle T_{\rm in}\rangle$ so closely. A natural possibility is that changing
$\langle T_{\rm in}\rangle$ shifts much of $P_M(T)$ with the overall thermal
scale. We test this directly in the \Appendix{} by expressing the
emission temperatures in units of $\langle T_{\rm in}\rangle$ and comparing
systems at the same scaled mass $M/\langle T_{\rm in}\rangle$. The resulting
scaled distributions become substantially more similar, although they do not
collapse exactly. At fixed scaled mass, identical distributions would give
$T_\beta\propto\langle T_{\rm in}\rangle$ directly from the definition of the
harmonic mean. The observed approximate rescaling therefore explains why
$T_\beta$, and through $T_{\rm loc}\simeq T_\beta$ the spectral inverse
slope, tracks $\langle T_{\rm in}\rangle$ so closely.

The rescaling is nevertheless incomplete. Fixed physical scales such as the
emission cutoff and momentum acceptance, together with the microscopic rate,
cooling history, and source composition, reshape $P_M(T)$ and produce
residual variation in $T_{\rm loc}/\langle T_{\rm in}\rangle$ even at matched
$M/\langle T_{\rm in}\rangle$. Moreover, the thermometer relation in
Fig.~\ref{fig:thermometer-mechanism}(a) is extracted over a fixed physical
mass window, for which $M/\langle T_{\rm in}\rangle$ itself changes along the
temperature scan. Temperature rescaling therefore provides the physical basis
for the strong thermometer correlation, but does not require either
proportionality or linearity. We do not identify a simple universal mechanism
that enforces the nearly linear
$T_{\llbar}$--$\langle T_{\rm in}\rangle$ relation; over the range studied
here, its near-linearity emerges from the full evolution and emission
response.

\section{Response to cooling and early source composition}

The harmonic-temperature interpretation makes clear which physical changes
modify the thermometer relation. Since the inverse-slope parameter is
controlled by the normalized temperature distribution $P_M(T)$ rather than
the total amount of radiation, an overall mass-independent rescaling of the
yield leaves $P_M(T)$ and $T_{\llbar}$ unchanged. By contrast, changes in the
evolution or source content that redistribute radiation among different
temperatures modify the harmonic mean and can change both the characteristic
value $T_{\llbar}^{*}$ and the response coefficient $B$.

\begin{figure}
\centering
\includegraphics[width=\linewidth]{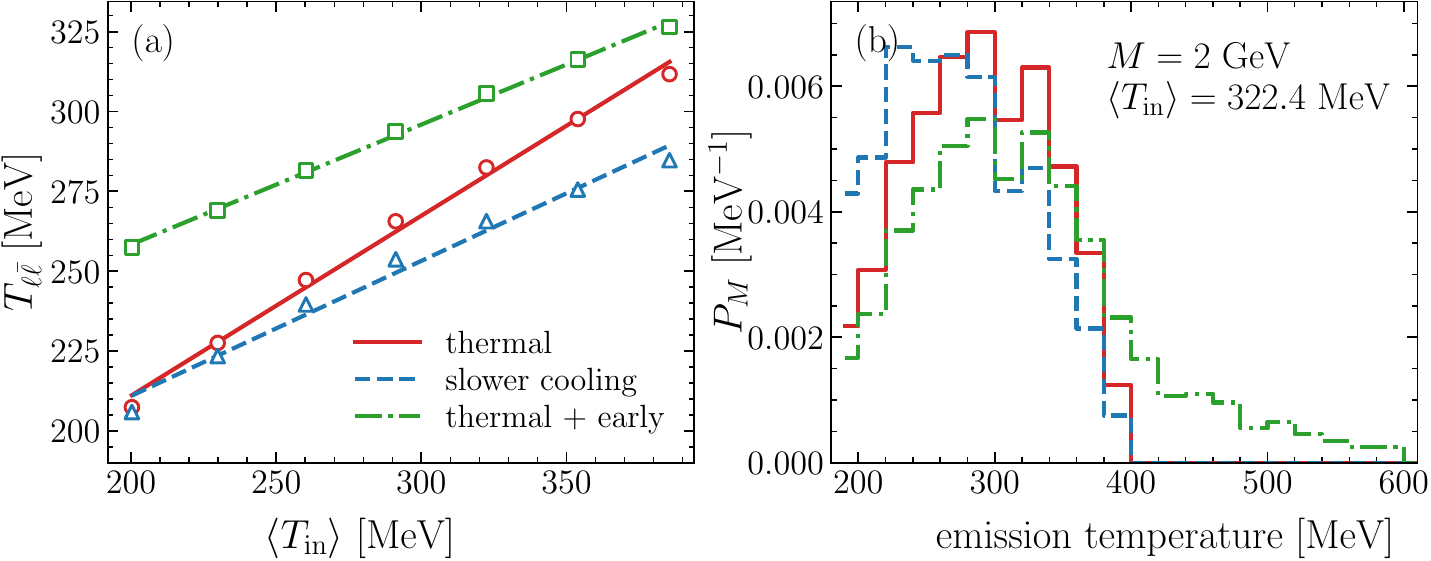}
\caption{(a) $T_{\llbar}$ versus the $e\gamma$-weighted thermal-stage initial
temperature for the baseline evolution, slower cooling, and thermal emission
supplemented by the early component. The corresponding response coefficients
are $B=0.564$, $0.423$, and $0.376$. (b) $P_M(T)$ at
$\langle T_{\rm in}\rangle=322.4~{\rm MeV}$ and $M=2~{\rm GeV}$, illustrating
how slower cooling and the early component redistribute the radiation toward
lower and higher temperatures, respectively.}
\label{fig:calibration-exposure}
\end{figure}

Figure~\ref{fig:calibration-exposure}(a) compares the baseline,
slower-cooling, and mixed-source scenarios introduced above. The same seven
sampled $\langle T_{\rm in}\rangle$ values are used in all cases, so the
differences in $B$ reflect changes in the physical response rather than
different sampled temperature ranges. Using Eq.~\eqref{eq:linear-response},
$B$ decreases from $0.564$ in the baseline calculation to $0.423$ for
slower cooling and $0.376$ when the early component is included. At the
common pivot $T_{\rm in}^{*}=291.9~{\rm MeV}$, the corresponding values of
$T_{\llbar}^{*}$ are $262.8$, $249.8$, and $292.9~{\rm MeV}$,
respectively. Cooling history and source composition therefore modify both
the characteristic inverse slope and its response to the thermal-stage
initial temperature.

Figure~\ref{fig:calibration-exposure}(b) illustrates the corresponding
redistribution of emission temperatures at the representative point
$\langle T_{\rm in}\rangle=322.4~{\rm MeV}$ and $M=2~{\rm GeV}$. Slower
cooling gives greater space-time weight to later, cooler radiation, shifting
$P_M(T)$ toward lower temperatures and thereby reducing $T_\beta(M)$ and
$T_{\llbar}$. The reduction of $B$ follows from how this redistribution
evolves along the temperature scan: as $\langle T_{\rm in}\rangle$
increases, the time required to cool to the emission cutoff grows more
rapidly for the slower-cooling evolution, so the additional cool contribution
becomes increasingly important. The resulting suppression of $T_{\llbar}$
is therefore stronger at the high-temperature end of the scan, reducing $B$.

The early component produces a different redistribution. At fixed
$\langle T_{\rm in}\rangle$, radiation from
$\tau_{\rm init}<\tau<\tau_0$ extends to effective temperatures above those
on the thermal initialization surface, shifting $P_M(T)$ upward and
increasing $T_{\llbar}$. Its effect on $B$ is controlled by how its relative
contribution changes along the temperature scan. For a mass window
$\mathcal W$, we define the integrated early fraction as
\begin{equation}
f_{\rm early}^{\mathcal W}
=
\frac{
\int_{\mathcal W} dM\,Y_{\rm early}(M)
}{
\int_{\mathcal W} dM\,[Y_{\rm th}(M)+Y_{\rm early}(M)]
}.
\label{eq:window-early-fraction}
\end{equation}
Here $Y_{\rm th}$ and $Y_{\rm early}$ denote the thermal and early
contributions to the total spectrum. In $1<M<3~{\rm GeV}$,
$f_{\rm early}^{\mathcal W}$ decreases from $45.6\%$ at the lowest
thermal-stage initial temperature to $8.7\%$ at the highest. In the present
construction, increasing $\langle T_{\rm in}\rangle$ enhances the thermal
contribution more rapidly than the early contribution. The early source
therefore raises $T_{\llbar}$ most strongly at low
$\langle T_{\rm in}\rangle$, flattening the thermometer relation and
reducing $B$. The direction and magnitude of this effect are not universal
properties of pre-equilibrium radiation, but depend on how the early source
evolves relative to the thermal medium. Throughout this comparison,
$\langle T_{\rm in}\rangle$ denotes the thermal-stage initial temperature at
$\tau_0$, not the largest effective temperature entering the mixed-source
emission history.

As a complementary normalization check, within the thermal baseline we vary
the transverse size $\sigma_0$ and rescale the initialization time $\tau_0$
while preserving the same temperature history in the corresponding scaled
coordinates. These changes strongly modify the space-time volume and total
dilepton yield but leave $P_M(T)$ and $T_{\llbar}$ unchanged. This confirms
within the controlled model that the thermometer relation is sensitive to how
radiation is distributed over temperature, rather than to its overall
normalization. This also provides a simple interpretation of part of the
robustness found in hydrodynamic calculations across centralities and beam
energies \cite{Churchill:2023zkk}: changes that mainly affect the system size,
lifetime, or overall radiation yield need not modify it unless they also
reshape $P_M(T)$.

\section{Mass-window leverage and source sensitivity}
The previous section shows that a single $T_{\llbar}$ need not uniquely
determine the thermal-stage initial temperature when an additional early
source contributes. The mass dependence of the same spectrum provides
additional leverage because the Boltzmann suppression becomes increasingly
selective at larger invariant mass. Harder mass windows suppress cooler
thermal radiation more strongly and can also change the relative contribution
of an additional source with a different spectral hardness. We examine these
effects by raising the lower fit boundary $M_{\min}$ while keeping
$M_{\max}=3~{\rm GeV}$ fixed.

\begin{figure}
\centering
\includegraphics[width=\linewidth]{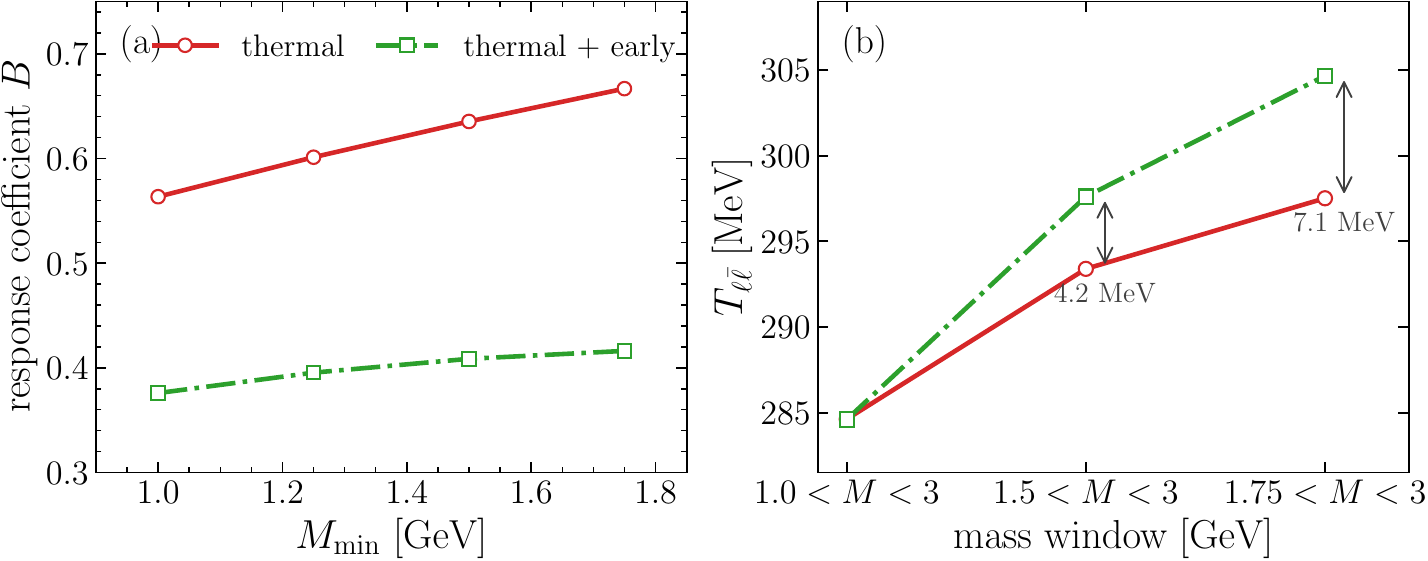}
\caption{(a) Response coefficient $B$ as the lower boundary of the mass
window is raised with $M_{\max}=3~{\rm GeV}$. The thermal response strengthens
at larger mass, whereas the increase is weaker when the harder early
component contributes. (b) Two systems with the same
$T_{\llbar}=284.6~{\rm MeV}$ in $1<M<3~{\rm GeV}$ develop inverse-slope
differences of $4.2~{\rm MeV}$ in $1.5<M<3~{\rm GeV}$ and
$7.1~{\rm MeV}$ in $1.75<M<3~{\rm GeV}$.}
\label{fig:mass-window}
\end{figure}

Using the response coefficient $B$ defined in
Eq.~\eqref{eq:linear-response}, Fig.~\ref{fig:mass-window}(a) shows that for
thermal emission $B$ increases from $0.564$ for $1<M<3~{\rm GeV}$ to
$0.601$, $0.635$, and $0.667$ as $M_{\min}$ is raised to $1.25$, $1.50$,
and $1.75~{\rm GeV}$, respectively. At larger invariant mass, cooler thermal
radiation is increasingly suppressed, so the spectrum is weighted more
strongly toward hotter regions of the thermal evolution. The extracted
$T_{\llbar}$ therefore responds more strongly to changes in
$\langle T_{\rm in}\rangle$ and hence gives a larger $B$. With the early
component included, however, $B$ increases much more weakly, from $0.376$ to
$0.395$, $0.409$, and $0.416$. This behavior reflects a competition between
two effects. Raising $M_{\min}$ strengthens the thermal response as above,
but it also increases the relative contribution of the harder early source.
As shown in Fig.~\ref{fig:calibration-exposure}(a), this source flattens the
thermometer relation in the present construction. Its increasing weight at
higher mass therefore counteracts part of the stronger thermal response,
leading to the weaker increase of $B$.

The additional information carried by the mass dependence can be illustrated
by constructing two systems with the same broad-window inverse-slope
parameter, $T_{\llbar}=284.6~{\rm MeV}$ in $1<M<3~{\rm GeV}$. A purely
thermal system requires $\langle T_{\rm in}\rangle=326.6~{\rm MeV}$, whereas
a thermal-plus-early system reaches the same $T_{\llbar}$ with a substantially
lower thermal-stage initial temperature,
$\langle T_{\rm in}\rangle=267.7~{\rm MeV}$. In the latter case, radiation
from $\tau<\tau_0$ extends to effective temperatures above those on the
thermal initialization surface, allowing the harder early contribution to
compensate for the colder thermal evolution. The two systems are therefore
degenerate when characterized only by their broad-window $T_{\llbar}$.

This compensation does not persist when the mass selection is changed. As
shown in Fig.~\ref{fig:mass-window}(b), the extracted $T_{\llbar}$ values
differ by $4.2~{\rm MeV}$ in $1.5<M<3~{\rm GeV}$ and by $7.1~{\rm MeV}$ in
$1.75<M<3~{\rm GeV}$. Using Eq.~\eqref{eq:window-early-fraction}, the early
fraction in the mixed system increases from $21.0\%$ in
$1<M<3~{\rm GeV}$ to $31.9\%$ and $38.4\%$ in the two harder windows. The
harder early spectrum is therefore progressively exposed as $M_{\min}$ is
raised, breaking the degeneracy present in the broad-window inverse-slope
parameter. Different mass windows thus provide complementary sensitivity to
the thermal temperature scale and the relative early-source contribution.
Measurements of $T_{\llbar}$ in multiple mass windows, when confronted with
realistic calculations of the thermal and early contributions, can therefore
help constrain source mixtures that remain degenerate in a single window.

\section{Summary and outlook}
We have identified the physical origin and principal limits of
intermediate-mass dilepton thermometry. The local inverse-slope parameter
follows, to percent-level accuracy, the emission-weighted harmonic mean of the
temperatures contributing to the spectrum. As the thermal-stage initial
temperature changes, much of $P_M(T)$ shifts with the overall thermal scale,
causing this harmonic temperature, and hence the spectral inverse slope, to
track $\langle T_{\rm in}\rangle$ closely. The rescaling is not exact:
fixed physical scales, cooling history, and source composition reshape
$P_M(T)$ and make the thermometer relation nonuniversal. Temperature
rescaling therefore explains why a strong correlation with
$\langle T_{\rm in}\rangle$ is natural, but does not by itself require the
nearly linear relation observed over the temperature range studied here.

This picture also clarifies which physical changes modify the thermometer
relation. Changes that only rescale the amount of radiation leave
$T_{\llbar}$ unchanged, whereas cooling history and source composition
redistribute radiation among different temperatures and modify both the
characteristic inverse slope and its response to
$\langle T_{\rm in}\rangle$. The IMR inverse-slope parameter is therefore a
quantitative but nonuniversal probe of the early thermal history.

The invariant-mass dependence provides additional information when more than
one emission source contributes. Harder mass windows preferentially select
hotter thermal radiation and can also increase the relative weight of a
harder early contribution. Consequently, source scenarios that are degenerate
in a broad-window $T_{\llbar}$ need not remain degenerate when the mass
selection is changed. Measurements of $T_{\llbar}$ in several invariant-mass
windows, when confronted with realistic calculations, can therefore provide
complementary constraints on the thermal evolution and early electromagnetic
source content. A quantitative extraction will ultimately require realistic
treatment of backgrounds, especially heavy flavor. Extending this framework
across the Beam Energy Scan and toward lower collision energies
\cite{Goes-Hirayama:2025lhs,Wu:2025iix,Bzdak:2019pkr,An:2021wof,
Du:2024wjm,NA60:2022sze}, with explicit treatment of the finite
nuclear-overlap heating stage and its nonequilibrium chemical evolution, will
test how robustly the thermometer mapping persists in these regimes and what
additional information its departures from universality may contain.

%% file: app.tex
\section*{Temperature rescaling and residual reshaping}
\label{sec:supp-scaling-response}

To make the discussion self-contained, we first recall the quantities used to
characterize the temperature content and local slope of the dilepton spectrum.
For a given invariant mass $M$, the normalized emission-temperature
distribution is
\begin{equation}
P_M(T;\langle T_{\rm in}\rangle)
\equiv
\frac{1}{\widetilde{Y}(M)}
\frac{d\widetilde{Y}(M)}{dT},
\qquad
\int dT\,P_M(T;\langle T_{\rm in}\rangle)=1,
\label{eq:supp-temperature-distribution}
\end{equation}
where $\widetilde{Y}(M)=M^{-3/2}Y(M)$. The corresponding harmonic emission
temperature and local inverse-slope parameter are
\begin{equation}
T_\beta(M)
\equiv
\left[
\int dT\,\frac{P_M(T)}{T}
\right]^{-1},
\qquad
T_{\rm loc}(M)
\equiv
-\left[
\frac{d\ln\widetilde{Y}(M)}{dM}
\right]^{-1}.
\label{eq:supp-harmonic-local}
\end{equation}
For the mixed-source calculation, the same notation is used for the effective
temperature entering the early contribution. As shown in the main text,
$T_{\rm loc}(M)$ follows $T_\beta(M)$ to percent-level accuracy throughout
the IMR for the scenarios considered here. Understanding why the spectral
inverse slope tracks $\langle T_{\rm in}\rangle$ therefore reduces to asking
how the temperatures sampled by the radiation change with the thermal-stage
initial temperature.

We test whether much of this change can be described as an overall rescaling
of $P_M(T)$. At fixed scaled mass, define
\begin{equation}
z\equiv M/\langle T_{\rm in}\rangle,
\qquad
\theta\equiv T/\langle T_{\rm in}\rangle.
\label{eq:supp-scaled-variables}
\end{equation}
For temperature point $i$, we evaluate the spectrum at
$M_i=z\langle T_{{\rm in},i}\rangle$ and introduce the normalized scaled
distribution
\begin{equation}
p_i(\theta;z)
\equiv
\langle T_{{\rm in},i}\rangle
P_{M_i}
\left(
T=\theta\langle T_{{\rm in},i}\rangle;
\langle T_{{\rm in},i}\rangle
\right),
\label{eq:supp-scaled-distribution}
\end{equation}
with $\int d\theta\,p_i(\theta;z)=1$.
This is an exact change of variables. The harmonic emission temperature then
satisfies
\begin{equation}
\frac{T_{\beta,i}(z)}
{\langle T_{{\rm in},i}\rangle}
=
\left[
\int d\theta\,
\frac{p_i(\theta;z)}{\theta}
\right]^{-1}.
\label{eq:supp-scaled-harmonic}
\end{equation}
Equation~\eqref{eq:supp-scaled-harmonic} makes the role of the rescaling
explicit. If $p_i(\theta;z)$ were identical for all systems at fixed $z$,
the right-hand side would be independent of $i$, giving
\begin{equation}
T_{\beta,i}(z)
=
C(z)\langle T_{{\rm in},i}\rangle ,
\label{eq:supp-perfect-scaling}
\end{equation}
where $C(z)$ is determined by the common scaled distribution. Perfect
temperature rescaling would therefore imply exact proportionality between
$T_\beta$ and $\langle T_{\rm in}\rangle$ at fixed $z$. Approximate
similarity of the scaled distributions naturally gives the strong, though
not exact, tracking observed in the calculation.

This fixed-$z$ result should not be confused with the thermometer relation in
the main text, which is extracted over a fixed physical mass window. Changing
$\langle T_{\rm in}\rangle$ there also changes
$z=M/\langle T_{\rm in}\rangle$, while $p_i(\theta;z)$ itself retains
residual system dependence. Temperature rescaling can therefore explain the
strong correlation without requiring the finite-window thermometer relation
to be exactly proportional or linear.

\begin{figure}[!t]
\centering
\includegraphics[width=\linewidth]
{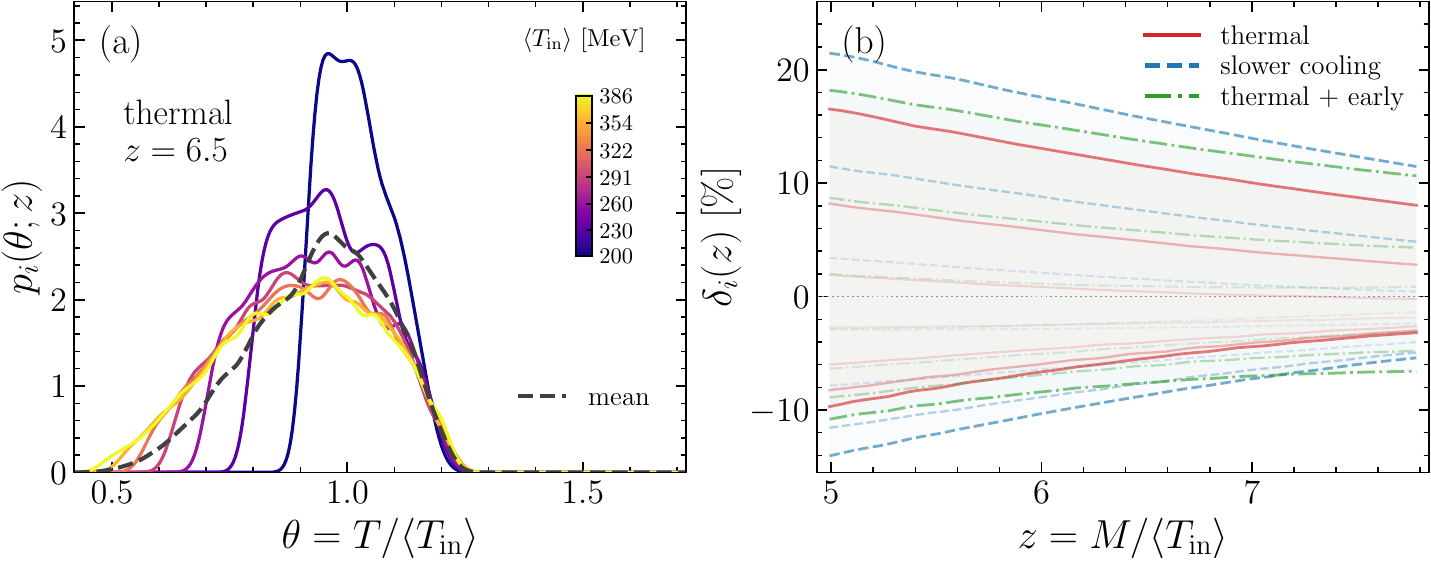}
\caption{%
Temperature rescaling and residual reshaping.
(a) Scaled emission-temperature distributions $p_i(\theta;z)$ for the seven
thermal-baseline calculations at
$z=M/\langle T_{\rm in}\rangle=6.5$, with
$\theta=T/\langle T_{\rm in}\rangle$. The dashed curve denotes their mean.
Rescaling by $\langle T_{\rm in}\rangle$ reduces the normalized pairwise
Wasserstein spread to $0.214$ of its value in physical temperature,
corresponding to a $79\%$ reduction, while residual shape differences remain.
(b) Fractional deviations of
$T_{\rm loc}/\langle T_{\rm in}\rangle$ from its scan mean at matched $z$.
The RMS deviations are $6.26\%$, $8.84\%$, and $7.40\%$ for the thermal
baseline, slower-cooling, and thermal-plus-early calculations, respectively.
All corresponding masses lie within $1\leq M\leq3~{\rm GeV}$.
}
\label{fig:supp-scaling-response}
\end{figure}

Figure~\ref{fig:supp-scaling-response}(a) tests the rescaling directly for the
thermal baseline at the representative value $z=6.5$. The seven distributions
become substantially more similar when temperature is expressed in units of
$\langle T_{\rm in}\rangle$. The full distribution is relevant rather than
only its peak, since $T_\beta$ is a harmonic mean and is also sensitive to
the lower-temperature part of the emission. To quantify the improvement, we
compare the mean pairwise Wasserstein distance before and after rescaling. For
a dimensionless comparison, the distance in physical temperature is
normalized by the mean $\langle T_{\rm in}\rangle$ of each pair. At $z=6.5$,
the remaining spread is $0.214$ of its value in physical temperature,
corresponding to a $79\%$ reduction. The same qualitative reduction is found
at the other tested values of $z$ and for the slower-cooling and
thermal-plus-early calculations.

The remaining shape differences have direct physical origins. In particular,
the fixed emission cutoff $T_{\min}=0.18~{\rm GeV}$ appears in scaled
coordinates at
\begin{equation}
\theta_{\min}
=
T_{\min}/\langle T_{\rm in}\rangle,
\label{eq:supp-scaled-cutoff}
\end{equation}
so its location changes along the temperature scan and becomes especially
important for the coolest systems. Slower cooling increases the relative
weight of radiation emitted near this fixed scale, while the early component
introduces additional scales through its effective-temperature range and
chemical evolution. The dominant temperature rescaling is therefore
accompanied by genuine reshaping of $P_M(T)$.

To determine whether this residual reshaping affects the spectral response,
Fig.~\ref{fig:supp-scaling-response}(b) compares the dimensionless local
response at matched $z$,
\begin{equation}
R_i(z)
\equiv
\frac{
T_{{\rm loc},i}
\left(z\langle T_{{\rm in},i}\rangle\right)
}{
\langle T_{{\rm in},i}\rangle
},
\qquad
\overline{R}(z)
\equiv
\frac{1}{N}\sum_i R_i(z).
\label{eq:supp-scaled-local-response}
\end{equation}
We plot its fractional deviation from the scan mean,
\begin{equation}
\delta_i(z)
\equiv
\frac{R_i(z)-\overline{R}(z)}{\overline{R}(z)}.
\label{eq:supp-matched-z-deviation}
\end{equation}
If the scaled distributions were identical, their harmonic temperatures would
coincide in units of $\langle T_{\rm in}\rangle$ and, through
$T_{\rm loc}\simeq T_\beta$, the scaled local responses would be nearly the
same. Figure~\ref{fig:supp-scaling-response}(b) instead shows appreciable
residual variation. Over the common interval
$4.99\lesssim z\lesssim7.78$, the RMS deviations are $6.26\%$, $8.84\%$,
and $7.40\%$ for the thermal baseline, slower-cooling, and
thermal-plus-early calculations, respectively. All corresponding masses
remain within $1\leq M\leq3~{\rm GeV}$, so no mass extrapolation is required.
The residual reshaping therefore has a non-negligible effect on the
dimensionless spectral response.

Together, the two panels clarify the role and limits of temperature
rescaling. The thermal-stage initial temperature sets a dominant overall
scale for the temperatures sampled by dilepton radiation, causing their
harmonic mean and hence the local inverse slope to track
$\langle T_{\rm in}\rangle$ closely. Fixed physical scales, cooling history,
and source composition nevertheless reshape $P_M(T)$ and make the response
nonuniversal. Temperature rescaling therefore explains why a strong
thermometer correlation is natural, but does not by itself explain the nearly
linear $T_{\llbar}$--$\langle T_{\rm in}\rangle$ relation observed in the
main text.

%% file: main.bbl
\begin{thebibliography}{52}
\bibitem [{fnl()}]{fnlinear}%
  \BibitemOpen
  \href@noop {} {}\bibinfo {note} {The previous hydrodynamic study fitted
  $\langle T_{\rm in}\rangle$ as a function of $T_{\llbar}$ for the purpose of
  temperature inference. Here we fit $T_{\llbar}$ as a function of $\langle
  T_{\rm in}\rangle$ to quantify the spectral response to controlled changes of
  the initial temperature.}\BibitemShut {Stop}%
\end{thebibliography}
